\documentclass[12pt]{book}

\usepackage[dvips]{graphicx,color}
\usepackage{makeidx,tsukuba}

\makeauthorindex
\makeindex

\begin{document}

\BookTitle{\itshape The 28th International Cosmic Ray Conference}
\CopyRight{\copyright 2003 by Universal Academy Press, Inc.}
\pagenumbering{arabic}

\chapter{
Mass Composition of the Primary Cosmic Rays
in the Energy Region $10^{14}\div 10^{20}$
eV in Anomalous Diffusion Model}

\author{%
%
%
A.A. Lagutin,$^1$ R.I. Raikin,$^1$ N.V. Stanovkina,$^1$ A.G. Tyumentsev,$^1$ A. Misaki$^2$ \\
{\it (1) Department of Theoretical Physics, Altai State University, Dimitrova str. 66,
Barnaul 656099, Russia\\
(2) Advenced Research Institute for Science and Engineering, Waseda University,
Ikubo 3-4-1, Shinjuku,Tokyo 169, Japan} }

\section*{Abstract}
We discuss the problem of cosmic ray mass composition variation
in wide energy region $10^{14}\div 10^{20}$ eV. The mass composition predicted
in the framework of recently developed anomalous diffusion model is
tested using the results of CORSIKA calculations
and experimental data on the depth of maximum of extensive
 air showers. We show that the model predictions for the mass
composition are consistent with observations of different experiments.

\section{Introduction}

The measurement of the mass composition of the primary cosmic rays (PCR),
especially with energies $\geq 10^{19}$ eV, is known to be one of the most
actual problems of modern astrophysics.
The reliable data about composition of the PCR are necessary
for solving the problems of the origin of ultrahigh energy cosmic ray and
also mechanisms  of their acceleration and propagation through the interstellar medium.

In recent works [1-5] the new interpretation of experimental data
 in the framework of galactic approach were offered under assumption that cosmic-ray
particles propagate through the fractal interstellar medium
(the anomalous diffusion model).
It was shown that the  "knee" ($E_0 \approx 3 \times 10^{15}$ eV) and the "ankle"
  ($E_0 \approx 8 \times 10^{18}$ eV) in PCR
  energy spectrum and also phenomenon of absence of
  Greisen-Zatsepin-Kuzmin cutoff [6,7] could be explaned
  in natural way in the anomalous diffusion model (see, figure 1).

At present stage, analysis and comparison of the experimental data on
extensive air showers (EAS)
 with results of EAS simulations is
 the only single effective method for the estimation of a
 PCR energy spectrum, composition and also for the test of hadronic interaction models at
 ultrahigh energies. One of the
 basic EAS characteristics, which can be used to determine
 the cosmic ray mass composition, is the
 depth of shower maximum  $t_{max}$.

In this paper the results of calculations  of $t_{max}$ of EAS,
simulated by CORSIKA(v.6.0) [8] in wide energy region for different primaries
are presented. The mass composition in energy region $10^{14}\div 10^{20}$ eV,
predicted in the framework of
 anomalous diffusion model, is compared
with experimental data of different arrays, which use various detection techniques:
 Casa-Blanca [9], Space-Vulcan [10], Hegra-Airobicc [11],
Fly's Eye [12], HiRes-MIA [13] and Yakutsk [14].

\begin{figure}[t]
 \begin{center}
\includegraphics[height=29.5pc, width=23.5pc, angle=-90]{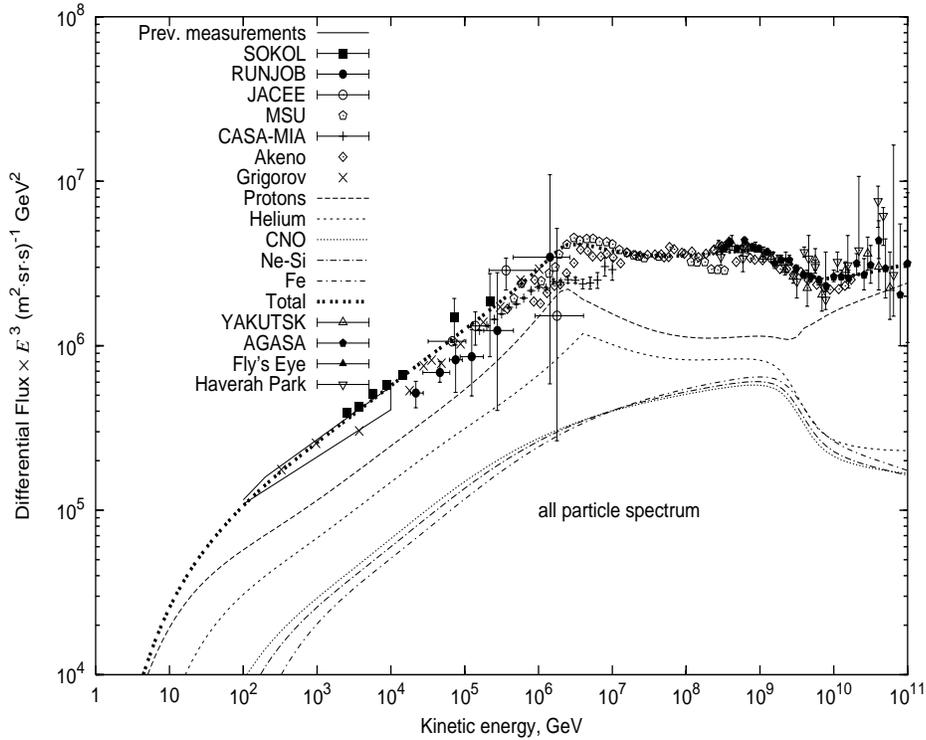}
 \end{center}
{\caption {The all particle spectrum calculated in
the anomalous diffusion model in comparison with the data
of different experiments [5].}} \hfill
\end{figure}

\section{Calculation and results}

 We used the quark-gluon string model with jets (QGSjet) which
provides the best overall
description of the experimental data about EAS [15]. The calculations have been made
for five groups of nuclei (p, He, CNO, Ne-Si, Fe), in the standard atmosphere.
For each primary particle
from 500 ($E_0 = 10^{14}$ eV) to 50 ($E_0 = 10^{20}$ eV)
events were simulated with thinning level $10^{-6}$.

\begin{figure}[t]
  \begin{center}
\includegraphics[height=25.5pc, width=27pc]{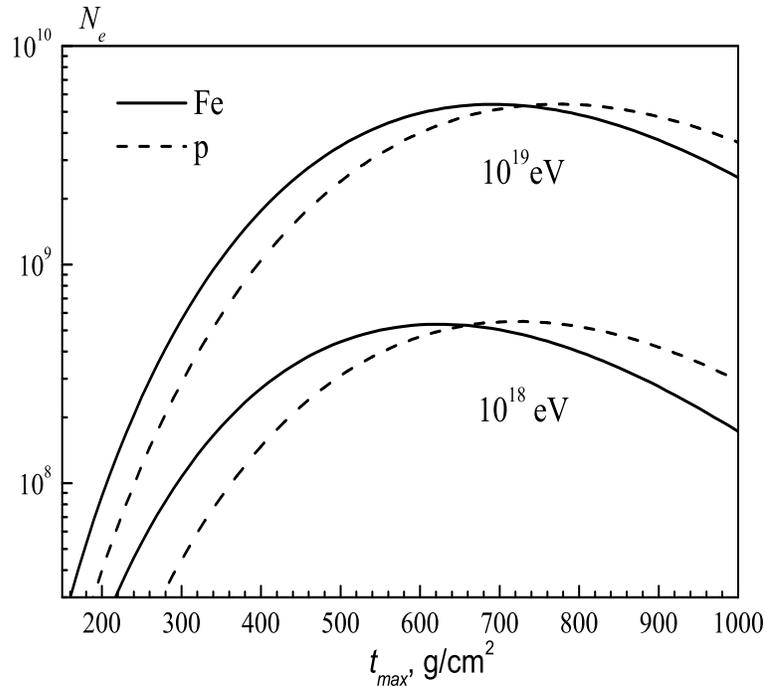}
  \end{center}
 \vspace{-0.5pc}
{\caption {Cascade curves for proton and iron
induced vertical showers at primary energies of $10^{18}$ and $10^{19}$ eV.}} \hfill
\end{figure}

\begin{table}
 \caption{Approximation coefficients of $t_{max}(E)$ (see (1)) for different primaries.}
\begin{center}
\begin{tabular}{l|ccc}
\hline type of nuclei & a & b & c \\
 \hline
p &-801.55 & 119.84 & -1.97 \\
He&-878.11&120.68 & -1.86 \\
CNO&-1131.84&142.84 & -2.38 \\
Ne-Si&-1164.74 & 141.26 & -2.26 \\
Fe&-1149.86 & 135.01 & -2.01 \\
\hline
\end{tabular}
\end{center}
\end{table}

In figure 2 we show cascade curves of electrons
for vertical showers induced by proton and iron of
 $10^{19}$ eV.  The depths of EAS maximum  were approximated by quadratic
expression of the form

\begin{equation}
t_{max} = a + b\cdot lg E + c\cdot (lg E)^2.
\end{equation}
 In table 1 we show the
approximation coefficients $a, b, c$ for different groups of nuclei.

\begin{table}[t]
 \caption{The mass composition of  primary
 cosmic rays in anomalous diffusion model.}
\begin{center}
\begin{tabular}{l|cccccr}
\hline
E, eV & $p_p$ & $p_{He}$ & $p_{CNO}$ & $p_{Ne-Si}$ & $p_{Fe}$ & $\langle A \rangle$ \\
\hline
$10^{14}$ & 0.43 & 0.25 & 0.12 & 0.11 & 0.09 & 10.91 \\
$3\cdot 10^{14}$ & 0.46 & 0.24 & 0.11 & 0.10 & 0.09 & 10.59  \\
$10^{15}$ & 0.51 & 0.23 & 0.09 & 0.09 & 0.08 & 9.49 \\
$3\cdot 10^{15}$ & 0.50 & 0.26 & 0.08 & 0.08 & 0.08 & 9.01 \\
$10^{16}$ & 0.41 & 0.26 & 0.11 & 0.11 & 0.11 & 11.55 \\
$3\cdot 10^{16}$ & 0.35 & 0.25 & 0.13 & 0.13 & 0.14 & 14.09 \\
$10^{17}$ & 0.31 & 0.23 & 0.15 & 0.15 & 0.16 & 15.84 \\
$3\cdot 10^{17}$ & 0.30 & 0.22 & 0.15 & 0.16 & 0.17 & 16.73\\
$10^{18}$ & 0.30 & 0.21 & 0.15 & 0.16 & 0.18 & 17.04 \\
$3\cdot 10^{18}$ & 0.39 & 0.18 & 0.13 & 0.14 & 0.16 & 15.25 \\
$10^{19}$ & 0.81 & 0.14 & 0.03 & 0.01 & 0.01 & 2.71\\
$3\cdot 10^{19}$ & 0.81 & 0.14 & 0.03 & 0.01 & 0.01 & 2.59 \\
$10^{20}$ & 0.81 & 0.14 & 0.03 & 0.01 & 0.01 & 2.59\\
\hline
\end{tabular}
\end{center}
\end{table}

For the mass composition, predicted in the framework of anomalous diffusion
model (table 2), the weighted depths of shower maximum  $t^{AD}_{max}(E) = \sum\limits_{i}
p_i\cdot t^{i}_{max}$, where $i$ denotes the group of nuclei,
were obtained according to CORSIKA calculations
  for five types of nuclei.
In figure 3 we present our results for the depth of maximum in
 proton- and iron-induced showers
and also $t^{AD}_{max}(E)$
in comparision with experimental data
 in the energy range $10^{14}\div 10^{20}$ eV.

\begin{figure}[t]
  \begin{center}
\includegraphics[height=26.5pc, width=29.5pc]{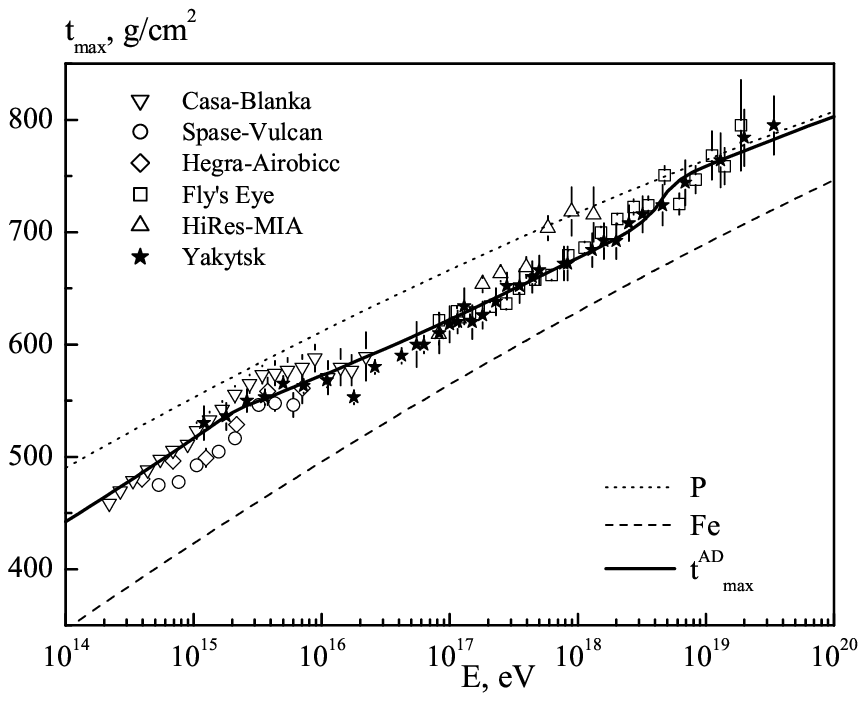}
  \end{center}
  \vspace{-0.5pc}
{\caption {Comparison of the weighted depth
of shower maximum in anomalous diffusion model with
experimental data [9-14].}}\hfill
\end{figure}

It must be emphasized that experimental data, which
consistently demonstrate the remarkable lightening of the primary composition
around the main features in energy spectrum
($E_0\approx 3\times 10^{15}$ eV and $E_0\approx 8\times 10^{18}$ eV),
 and our results obtained in the
framwork of anomalous diffusion model
are in satisfactory agreement in the whole considered energy range.

\section{Conclusions}

We have considered the mass composition
variation of primary cosmic rays at energy $10^{14}\div 10^{20}$ eV. CORSIKA/QGSjet
simulations of extensive air showers for primary p, He, CNO, Ne-Si, Fe in a wide
energy range were performed. The depth of shower maximum were estimated for
 mass composition predicted in the framework of
anomalous diffusion model. It was shown
that the experimentally observed
 lightening of primary composition in the "knee" and the "ankle" regions,
 is the consequence of anomalous diffusion of cosmic rays in
fractal interstellar medium.

{\it This work was supported UR grant 02.01.014.}

\section{References}
\re
1.\ Lagutin A.A., Nikulin Yu.A., Uchaikin V.V.\
2001, Nucl. Phys.B (Proc. Suppl.), v.97, 267.
\re
2.\ Lagutin A.A.\ 2001, Problems of atomic science and technology, N6(2), 214.
\re
3.\ Lagutin A.A., Strelnikov D.V., and  Tyumentsev A.G.\ 2001,
Proc.of the 27th ICRC, v.5, 1896.
\re
4.\ Lagutin A.A., Uchaikin V.V.\ 2003, Nucl. Instr. and Meth. in Phys. Res.,
v. B201, 212.
\re
5.\ Lagutin A.A., Tyumentsev A.G. 2003, Izv. RAN, ser Fiz., v.67, N4.
\re
6.\ Greisen K.\ 1966, Phys. Rev. Letters, v.16, 748.
\re
7.\ Zatsepin G.T., Kuzmin V.A.\ 1966, JETP Letters, v.4, 78.
\re
8.\ Heck D., Knapp J.\ Extensive Air Shower Simulation with CORSIKA:
A User's Guide (Version 6.00 from December 13, 2000)
 - Kernforschungszentrum Karlsruhe GmbH. 2000.
\re
9.\ Fowler J.W., Fortson L.F., Jui C.C.H. et al.\  2000,
 Preprint astro-ph/0003190 v2, 25.
\re
10.\ Dickinson J.E., Gill J.R., Hart S.P. et al.\ 1999, Proc.of the 26th ICRC,
 3, 136.
\re
11.\ Arqueros F., Barrio J.A., Bernlohr et al.\ arXiv: astro - ph/9908202.
\re
12.\ Bird D. et al.\ 1993, Phys. Rev. Lett.,  v.71, 3401.
\re
13.\ Abu-Zayyad T., Belov K., Bird D.J. et al.\ 2000,
 Phys. Rev. Lett., v.84, N19, 4276.
\re
14.\ Knurenko S., Kolosov V., Petrov Z., et al.\ 2001, Proc.of the 27th ICRC, v.1, 177.
\re
15.\ Kalmukov N.N., Ostapchenko S.S.\ 1993, Yad. Fiz, v.56, N3, 105.

\endofpaper
\end{document}